\documentclass[conference]{IEEEtran}
\IEEEoverridecommandlockouts
\usepackage{cite}
\usepackage{amsmath,amssymb,amsfonts}
\usepackage{algorithmic}
\usepackage{graphicx}
\usepackage{textcomp}
\usepackage{url}
\usepackage{breqn}
\usepackage{upgreek}
\usepackage{soul}
\usepackage[super]{nth}
\usepackage[letterpaper, left=0.7in, right=0.7in, bottom=0.95in, top=0.7in]{geometry}
\usepackage{dsfont}
\usepackage[belowskip=-12pt,aboveskip=0pt]{caption}
\usepackage[font=small,labelfont=bf]{caption}
\UseRawInputEncoding
\begin{document}

\title{Neural Network Architectures for Location Estimation in the Internet of Things\\
}

\title{{Neural Network Architectures for Location Estimation in the Internet of Things}}
\author{
\IEEEauthorblockN{Ullah Ihsan$^1$, Robert Malaney$^1$, and Shihao Yan$^2$}
\IEEEauthorblockA{$^1$School of Electrical Engineering  \& Telecommunications,
University of New South Wales,
Sydney, NSW 2052, Australia \\
$^2$School of Engineering,
Macquarie University,
Sydney, NSW 2109, Australia}}

\maketitle

\begin{abstract}
Artificial Intelligence (AI) solutions for wireless location estimation are likely to prevail in many real-world scenarios. In this work, we demonstrate for the first time how the Cramer-Rao upper bound on localization accuracy can facilitate efficient neural-network solutions for wireless location estimation. In particular, we demonstrate how the number of neurons for the network can be intelligently chosen, leading to AI location solutions that are not time-consuming to run and less likely to be plagued by over-fitting. Experimental verification of our approach is provided. Our new algorithms are directly applicable to location estimates in many scenarios including the Internet of Things, and vehicular networks where vehicular GPS coordinates are unreliable or need verifying. Our work represents the first successful AI solution for a communication problem whose neural-network design is based on fundamental information-theoretic constructs. We anticipate our approach will be useful for a wide range of communication problems beyond location estimation.
\end{abstract}

\section{Introduction}\label{introduction}

Several traditional localization algorithms have been developed to estimate the location of a vehicle in the past, e.g., ~\cite{lee2009rf,golestan2012vehicle,alam2012relative,hoang2016cooperative,cruz2017neighbor}. The practical limitations with these algorithms may range from limited functionalities to a complete failure as the surrounding environment changes. Therefore, we need localization algorithms that are practically deployable, smart enough to adapt to the environmental changes, and that are realistic.

To address the challenges that traditional localization algorithms face, researchers in recent times have incorporated numerous neural-network and machine-learning algorithms for positioning users/devices~\cite{chen2017confi,kumar2018feed}. While neural-networks have been able to address the shallow learning capabilities of the classic machine-learning algorithms, a key question about neural-networks, that is yet to be answered, is how to design their internal architecture, i.e., the number of chosen hidden layers, the choice of activation function in the hidden layer(s), and the number of neurons in each hidden layer. While the research community follows hyperparameters search mechanisms to finalize the architecture for neural-network frameworks at large~\cite{bergstra2012random}, numerous guidelines have also been provided in the recent literature to formulate an optimal neural-network architecture~\cite{Goodfellow2016,heaton2015aifh}. However, to-date there has been no concrete solution on how to pre-determine a neural-network architecture for a given problem.

In this work, we develop a feedforward neural-network framework for location estimation and formulate an insight into its architecture. We use received signal strength (RSS) of the vehicles' transmitted signals measured at multiple static road side units (RSUs). Through analysis based on the Cramer-Rao upper Bound (CRB) on the location accuracy of a vehicle, we  identify an architecture for a neural-network-based location estimation framework (NNLEF). Detailed numerical analysis confirms our analysis.

Although the concepts discussed here are in the context of vehicular adhoc networks (VANETs), they are widely applicable to a range of location-centric applications within the domain of internet of things.  Beyond the contribution stated above we summarize our additional contributions thus:
\begin{enumerate}
\item {We derive a value for the number of neurons needed in the hidden layer.}
\item {Through simulated data, we show how the NNLEF (with the adopted architecture) outperforms when compared to other NNLEFs (that follow random architectures).}
\item {We further show how the NNLEF (with the adopted architecture) performs more efficiently when compared to a traditional RSS-based algorithm.}
\item {Finally, we  experimentally validate our recommended architecture.}
\end{enumerate}
The remainder of this paper is organized as follows. Section~\ref{SM} details the system model and the derivation of the CRB on location accuracy. Section~\ref{LE} presents the NNLEF. Section~\ref{NR} provides numerical results based on simulated and experimental data, and Section~\ref{Conclusion} concludes this paper.

\section{System Model and RSS Location Estimation}\label{SM}

We consider the following system model in our work:
\begin{enumerate}
\item {The true location of a random vehicle (which is unknown to the framework) is denoted by $\textbf{x}_t=[x_0,y_0]$.}
\item {The framework has $N$ number of RSUs with publicly known locations. The true location of the \textit{i}-th RSU is $\textbf{x}_i=[x_i,y_i]$ where $i=1,2,...,N$.}
\item {All the RSUs are in the transmission range of the randomly located vehicle and independently measure RSS (all RSS in dBm) of the transmitted signal (from the random vehicle) every second. We adopt a log-normal shadowing model for the RSS observations. The measured RSS at the \textit{i}-th RSU, i.e., $r_i$, is given as
\begin{equation*}
r_i[dBm]\,\mathrm{=}\,P_T[dBm]\,\mathrm{-}\,PL_{d_i}[dB],\\
\end{equation*}
where $P_T$ is the transmit power of the vehicle and $PL_{d_i}$ (the path loss at a distance $d_i$) is given by
\begin{multline*}
PL_{d_i}[dB]\,\mathrm{=}\,PL_{d_o}\mathrm{+}\,10\,\upgamma\,\log_{10}\Big(\frac{d_i}{d_0}\Big)+X_{\upvarsigma_{db}}, \\
    (i=1,2,\dots ,N),
\end{multline*}
  where $PL_{d_0}$ is the reference path loss at a reference distance $d_0$, $\upgamma$ is the path loss exponent, $d_i$ is the vehicle-RSU$_i$ distance $(d_i > d_0)$ given by $d_i=\sqrt{{(x_i-x_0)}^2+{(y_i-y_0)}^2}$, and $X_{\upvarsigma_{db}}$ is a zero mean normal random variable with variance $\upvarsigma_{db}^{2}$ representing the shadowing noise. The RSS measurements made by the $N$ RSUs are independent of each other. They collectively form an RSS vector given by $\textbf{r}={[r}_1,\ r_2,\dots ,{r_N]}$.}
\item {We choose one of the $N$ RSUs as the processing center (PC). The PC accumulates its RSS measurements with the regularly collected RSS measurements from all surrounding RSUs. The PC further processes these measurements to estimate the location of the random vehicle. The estimated location of the vehicle is denoted by $\hat{\textbf{x}}_e=[\hat{x}_e,\hat{y}_e]$.}
\end{enumerate}

\subsection{Cramer-Rao Upper Bound Derivation}\label{CRB}

We now derive the CRB on the location accuracy of a vehicle. For a random transmitting vehicle, whose location $\textbf{x}_t$ is unknown, and whose RSS is measured at $N$ RSUs, the distribution of the RSS takes the form (with few constant elements ignored)

\begin{equation*}
-\ln\!f_{r_i|\textbf{x}_t\textbf{x}_i} = \frac{[r_i+\upgamma\,(\frac{10}{\ln{10}})\,\ln\!\,(\frac{d_i}{d_0})]^2}{2\upvarsigma_{db}^{2}}.
\end{equation*}
The covariance matrix $\mathcal{C}$ (related to the position) can be written as the inverse of the Fisher information matrix, $\mathcal{F}$, i.e., $\mathcal{C}=\frac{1}{\mathcal{F}}$. The elements of $\mathcal{F}$ are given by
\begin{align*}\label{fisher}
\mathcal{F} =
\begin{bmatrix}
\mathcal{I}_{xx} & \mathcal{I}_{xy} \\
\mathcal{I}_{yx} & \mathcal{I}_{yy} \\
\end{bmatrix},
\end{align*}
where
\begin{equation*}\label{fisher_11}
\mathcal{I}_{xx} = -\mathds{E}\bigg[\frac{\partial^2}{\partial x_i\,\partial x_i}\big(\ln\!f_{r_i|\textbf{x}_t\textbf{x}_i}\big)\bigg],
\end{equation*}
\begin{equation*}\label{fisher_12}
\mathcal{I}_{xy} = -\mathds{E}\bigg[\frac{\partial^2}{\partial x_i\,\partial y_i}\big(\ln\!f_{r_i|\textbf{x}_t\textbf{x}_i}\big)\bigg],
\end{equation*}
\begin{equation*}\label{fisher_21}
\mathcal{I}_{yx} = -\mathds{E}\bigg[\frac{\partial^2}{\partial y_i\,\partial x_i}\big(\ln\!f_{r_i|\textbf{x}_t\textbf{x}_i}\big)\bigg],
\end{equation*}
\begin{equation*}\label{fisher_22}
\mathcal{I}_{xx} = -\mathds{E}\bigg[\frac{\partial^2}{\partial y_i\,\partial y_i}\big(\ln\!f_{r_i|\textbf{x}_t\textbf{x}_i}\big)\bigg],
\end{equation*}
where $\mathds{E}$ denotes the expectation operation. The expressions in brackets are given as,
\begin{align*}
& \frac{\partial^2}{\partial x_i\,\partial x_i}\,\,=\,\, \frac{a\,\upgamma}{\upvarsigma_{db}^{2}}
  \sum_{i=1}^{N}\,\frac{1}{d_i^2}\bigg[\frac{a\,\upgamma\,(x_i\,-\,x_0)^2}{d_i^2} + \\
& \phantom{{}={}} \phantom{{}={}} \phantom{{}={}} \phantom{{}={}} \phantom{{}={}} \begin{aligned}[t] \Big(r_i\,+\,a\,\upgamma\,\ln{(d_i)}\Big)\Big(1\,-\,\frac{2\,(x_i\,-\,x_0)^2}{d_i^2}\Big) \bigg],
\end{aligned}
\end{align*}
\begin{align*}
 &   \frac{\partial^2}{\partial y_i\,\partial y_i}\,\,=\,\,\frac{a\,\upgamma}{\upvarsigma_{db}^{2}}
    \sum_{i=1}^{N}\,\frac{1}{d_i^2}\bigg[\frac{a\,\upgamma\,(y_i\,-\,y_0)^2}{d_i^2} + \\
& \phantom{{}={}} \phantom{{}={}} \phantom{{}={}} \phantom{{}={}} \phantom{{}={}} \begin{aligned}[t] \Big(r_i\,+\,a\,\upgamma\,\ln{(d_i)}\Big)\Big(1\,-\,\frac{2\,(y_i\,-\,y_0)^2}{d_i^2}\Big) \bigg],
\end{aligned}
\end{align*}
\begin{align*}
 &   \frac{\partial^2}{\partial x_i\,\partial y_i}\,\,=\,\,\frac{\partial^2}{\partial y_i\,\partial x_i}\,\,=\,\,
    \frac{a\,\upgamma}{\upvarsigma_{db}^{2}}\sum_{i=1}^{N}\,\frac{(x_i\,-\,x_0)\,(y_i\,-\,y_0)}{d_i^4}\, \\
& \phantom{{}={}} \phantom{{}={}} \phantom{{}={}} \phantom{{}={}} \phantom{{}={}} \phantom{{}={}} \phantom{{}={}} \phantom{{}={}} \begin{aligned}[t] \bigg[a\,\upgamma\,-\,2\,\Big(r_i\,+\,a\,\upgamma\,\ln{(d_i)}\Big)\bigg],
\end{aligned}
\end{align*}
where $a=\frac{10}{\ln{(10)}}$.

We consider $d_0=1m$ and extract the final expressions for the elements of $\mathcal{F}$ as,
\begin{equation*}
\mathcal{I}_{xx}\,=\,\frac{a^2\upgamma^2}{\upvarsigma_{db}^{2}}
    \sum_{i=1}^{N}\frac{(x_i-x_0)^2}{d_i^4},\,\,
\mathcal{I}_{yy}\,=\,\frac{a^2\upgamma^2}{\upvarsigma_{db}^{2}}
    \sum_{i=1}^{N}\frac{(y_i-y_0)^2}{d_i^4},
\end{equation*}
\begin{equation*}
\mathcal{I}_{xy}\,=\,\mathcal{I}_{yx}\,=\,\frac{a^2\,\upgamma^2}{\upvarsigma_{db}^{2}}
    \sum_{i=1}^{N}\,\frac{(x_i-x_0)\,(y_i-y_0)}{d_i^4}.
\end{equation*}
The final expression for CRB can be written as, $\rho^2=\rho_{xx}^2 + \rho_{yy}^2$. Here $\rho_{xx}^2$ and $\rho_{yy}^2$ are the diagonal elements of $\mathcal{C}$, and $\sqrt{\rho^2}$ equals the standard deviation of the CRB.
\begin{figure}[t!]
\centering
\includegraphics[width=0.5\textwidth]{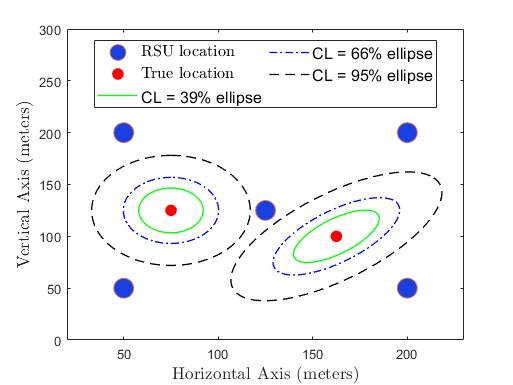}
\caption{Ellipses with confidence levels (CLs) of 39\%  (in solid green), 66\% (in dashed blue), and 95\% (in dashed black) are plotted for 2 sample test points. The test points in red form the origin of the respective ellipses.\label{Figure1}}
\end{figure}

To draw the ellipses representing the CRB on location accuracy in terms of confidence levels (CLs), the rotation matrix $\mathcal{R}$ is required, and is given by
\begin{align*}
    \mathcal{R} =
    \begin{bmatrix}
    \cos(\uptheta) & -\sin(\uptheta) \\
    \sin(\uptheta) & \,\,\,\,\,\cos(\uptheta) \\
    \end{bmatrix},
\end{align*}
where $\uptheta=tan^{-1}(\lambda_1/\lambda_2)$, while $\lambda_1$ and $\lambda_2$ are the eigenvectors of $\mathcal{C}$. The range of $\uptheta$ is in between $0$ and $2\pi$. The probability of a vehicle's location (returned by a positioning system) lying within a CL ellipse is given below \cite{malaney2007securing}
\begin{align*}
P_{in}=1-e^{-\frac{K}{2}},
\end{align*}
where $K$ is a constant that sets the scaling of the confidence ellipse.

Using our derivation, we draw in Fig.~\ref{Figure1} ellipses with different CLs for 2 sample test points. The locations of the RSUs are (50m, 50m), (50m, 200m), (125m, 125m), (200m, 50m), and (200m, 200m). The value of $\sigma_{db}$ is fixed at 5dB, and the path loss exponent equals 3.

\section{Neural-network based Location Estimation}\label{LE}

This section highlights the adopted NNLEF's performance in relation to estimating a vehicle's location. A feedforward neural-network forms the basis of this framework. This framework utilizes the measured RSS (influenced by the shadowing noise) at multiple RSUs. A feedforward network is a special type of neural-network that is known to manipulate and learn from the physical layer properties of the vehicles' transmitted signals~\cite{ihsan2019machine,ihsan2019location,ihsanGC2019}. Neural-network frameworks with a single, or multiple hidden layers, have the capability to converge to a continuous target function. However, a single hidden layer neural-network framework has a more flexible learning rate and converges faster to the target function when compared to a multiple hidden layer neural-network framework~\cite{nakama2011comparisons}. The architecture for the neural-network framework in this work is thus limited to a single hidden layer.

We next focus on developing an intuition into the performance of the NNLEF with changing activation functions and with a varying number of neurons in the hidden layer. It is evident from the expressions of a logistic sigmoid activation function, i.e., $a(x)=(1+e^{-x})^{-1}$, and a tangent sigmoid activation function, i.e., $a(x)=\frac{1-e^{-2x}}{1+e^{-2x}}$, that the gradient for both these activation functions at absolute high values is approximately zero. With complex data at the input, this phenomenon can minimize learning for the NNLEF. On the other hand, a steady gradient for the ReLu activation function, $a(x)=max[0,x]$, keeps the framework's learning consistent in the region where $x>0$ and there is a possibility of faster learning for the NNLEF~\cite{krizhevsky2012imagenet,bircanouglu2018comparison}. We have investigated and found that different transfer functions in the hidden layer of the NNLEF produce comparable results. In order to accommodate for future complex channel environments we take into account ReLu as the choice of transfer function for the hidden layer of the NNLEF.

\begin{figure}[t!]
\centering
\includegraphics[width=0.45\textwidth]{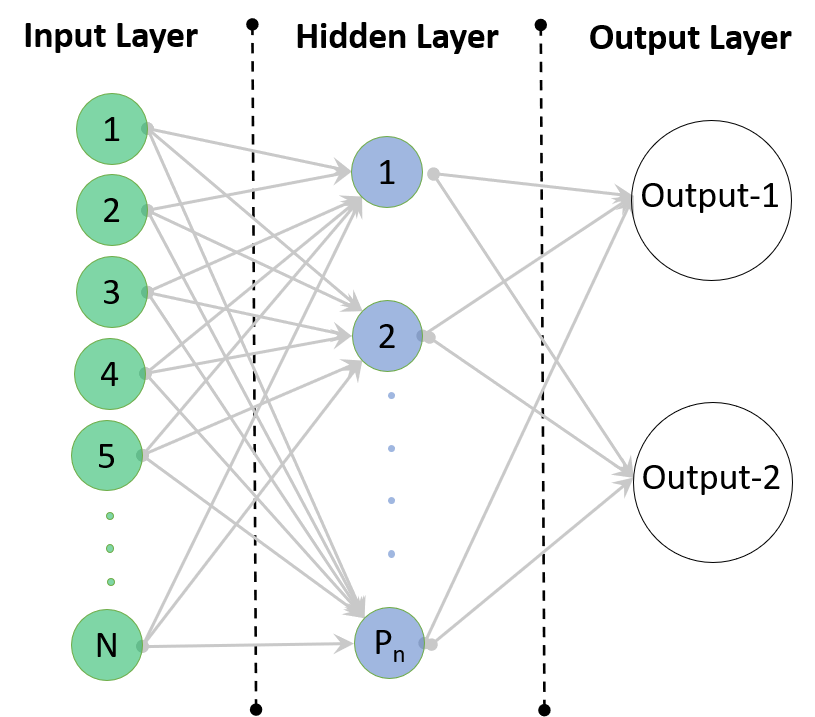}
\caption{A schematic of the feedforward neural network adopted for the NNLEF in this work. The number of inputs is set to $N$. The hidden layer has $P_n$ neurons. The number of outputs in the output layer is set to 2, i.e., the dimension of a pair (of coordinates).\label{Figure2}}
\end{figure}
The input to the NNLEF comprise $\textbf{r}$. The number of outputs is set to 2 (the location coordinates). A schematic of the NNLEF in this work is shown in Fig.~\ref{Figure2}. The number of neurons in the hidden layer, $P_n$, is our focal point in this study. The text in the following paragraphs will provide an insight into a recommendation for $P_n$. This recommendation for $P_n$ will allow for a promising performance of the NNLEF. Each neuron in the hidden layer of the NNLEF partially contributes towards the performance of the NNLEF. A very small $P_n$ is likely not sufficient to extract the hidden features/patterns in the input data. A reasonable $P_n$ is therefore required for the NNLEF to perform efficiently. Increasing $P_n$ is expected to produce good results (Figs.~\ref{Figure3} and \ref{Figure6} highlights this phenomenon) but is not, in general, advised as this only adds to the framework's overhead. That is, increasing $P_n$ can improve the NNLEF's performance marginally, but at the cost of an unnecessary increase in the NNLEF's overhead (e.g., the number of parameters, computational time, and memory resources). Additionally, with a very high $P_n$, the NNLEF can lead to an over-fitting problem especially in conjunction when too much training data is supplied under one specific channel condition.

Two critical questions are faced in the design of any neural network algorithm. One is: How much training of the algorithm should occur? A second is: What should the architecture be (how many neurons)?  Consider a channel that is perfectly described by some model, and this is used for training purposes. If we were to train a neural network under this model with unconstrained training samples (RSS values and all location information) and unconstrained $P_n$ (the number of neurons to be placed  in the hidden layer) we would be identifying perfectly, in effect, the function that describes the model's distance \emph{vs}. RSS relation. Given this perfect channel identification any use of that network to determine an unknown location from noisy RSS values should achieve a location error at the CRB\footnote{However, in  real-world scenarios we could expect the location accuracy of such a network be substantially less than the CRB. The reason for this is that, in general, the real-world channel will never be exactly the training model.}.

In estimating $P_n$, we would like to ensure that any loss in the network performance (distance accuracy) introduced by constraining that number is not too large. As we now show, the Universal Approximation Theorem (UAT) \cite{hornik1989multilayer} can allow insight into that. Loosely speaking, the UAT states that there always exists a neural network that can approximate any input function, $f(x)$, with any output function, $g(x)$, to any arbitrary accuracy $\varepsilon$, i.e.  $\left| {g(x) - f(x)} \right| < \varepsilon$. In our case $x$ $=$ RSS and $f(x)$ maps to the distance, $d$, between transmitter and receiver.

To make progress let us consider a single hidden layer neural network with transfer functions of the sigmoid form (our result will be independent of this choice). A neuron is modelled by $\zeta (\omega x + b)$, where $\zeta (z) \equiv 1/(1 + {e^{ - z}})$. It is straightforward to show that values of $\omega$ and $b$ can be chosen to `force' the transfer function into a step form where the step occurs at $-b/\omega$. Further, by using $P_n$ transfer functions ($P_n$ neurons) connected in a single layer network, it is straightforward to show  that a series of rectangles can be formed at the output \cite{nielson}. That is, you can create a network that can model any input function $f(x)$ as an output function $g(x)$ consisting of a series of $P_n$ rectangles. If we simplify this further and make all the rectangles of equal width, we can easily determine $P_n$ such that $\left| {g(x) - f(x)} \right| < \varepsilon$, where $\varepsilon$ now represents the standard deviation in the distance difference between the two functions. The CRB on the distance estimate for log-normal shadowing is $\ln (10){\sigma _{dB}}d/(10n)$. Therefore, a good estimate of our required $P_n$ would be one that ensures $\ln (10){\sigma _{dB}}d/(10n) >  \varepsilon$.
Carrying out the calculation detailed above, taking a typical distance scale of order 100m we find the following:
Adopting $n=3$ and $\sigma_{db} =3$ we find that $P_n=12$;
adopting $n=3$ and $\sigma_{db} =5$ we find we find that $P_n=9$;
and adopting $n=3$ and $\sigma_{db} =8$ we find we find that $P_n=7$;
This indicates $P_n$ in the range of 7-12 would be useful for the type of channels we investigate here.

\section{Numerical Results}\label{NR}

\subsection{Analysis Using Simulated Data}\label{SD}

We now present our numerical results by taking into account simulated data. The focus area is of size 200m$\times$200m. 5 RSUs are installed at (0m, 0m), (0m, 200m), (100m, 100m), (200m, 0m), and (200m, 200m). The focus area resembles a cross section of an expressway. The horizontal and vertical axes are partitioned into equidistant divisions. The RSUs measure RSS from the cross section of the divisions on both the axes at a frequency of 1Hz. The RSS measurements are under the influence of random shadowing noise. To mimic reality and to accommodate for the unique location of each RSU, this noise element is extracted from a random Gaussian distribution with a fixed $\sigma_{db}$. This means that at any given instant, the RSS measurements from a cross section of the divisions on all the RSUs will have unique and independent shadowing noise elements included in them. A value of 5dB is taken into account for $\sigma_{db}$. The path loss exponent is set to 3. After the RSS measurement campaign, the RSS database is randomized and further divided into two sets; a test set (with nearly 10\% of the database samples), and a training set (with the remaining database samples). The training set has the horizontal and vertical coordinates for the cross section of the divisions, while the test set has no such information included.

\begin{figure}[t!]
\centering
\includegraphics[width=0.5\textwidth]{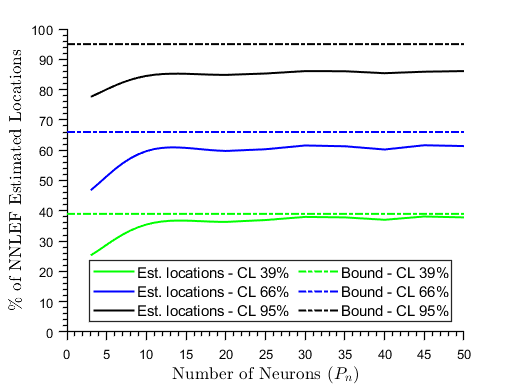}
\caption{The percentage of estimated test locations by NNLEFs with a changing $P_n$. The number of RSUs is 5 and the value of $\sigma_{db}$ is set to 5dB. The horizontal axis shows the number of neurons in the hidden layer, $P_n$, for different NNLEFs. The solid green, blue, and black lines represent the percentage of estimated locations for vehicles in ellipses with CLs 39\%, 66\%, and 95\%, respectively. The dashed lines indicate the Cramer-Rao upper bounds on location accuracy for the corresponding confidence ellipses. We see that the performance for the NNLEF becomes nearly steady as $P_n$ equals or exceeds 8.\label{Figure3}}
\end{figure}

In Fig.~\ref{Figure3} we study the performance of the NNLEFs with changing $P_n$. For uniformity, we use a ReLu activation function in the hidden layer of all the NNLEFs. Moreover, all the other neural-network training parameters are kept the same. We train all the NNLEFs with the same training set data. Once the training concludes, we subject the NNLEFs to estimate locations for the  vehicles in the test set. To analyze the performance, we draw confidence ellipses with different CLs and plot the estimated locations by each NNLEF. To determine whether an NNLEF's estimated location is within or outside a particular confidence ellipse, we use the equation below
\begin{dmath*}
    \frac{[\cos(\uptheta)(\hat{x}_e-x_c)+\sin(\uptheta)(\hat{y}_e-y_c)]^2}{\mathnormal{l}_{maj}}\,+\,\frac{[\sin(\uptheta)(\hat{x}_e-x_c)-\cos(\uptheta)(\hat{y}_e-y_c)]^2}{\mathnormal{l}_{min}}\,\,\leqslant1,
    \label{ellipse_eq}
\end{dmath*}
where $(x_c,y_c)$ is the center of the ellipse, $\mathnormal{l}_{maj}$ is the length of the semi-major axis, and $\mathnormal{l}_{min}$ is the length of the semi-minor axis for a particular confidence ellipse. In Fig.~\ref{Figure3} we plot the percentage of the NNLEFs' estimated test locations in each confidence ellipse (on the y-axis) against $P_n$ (on the x-axis). The changing $P_n$ on x-axis corresponds to different NNLEFs. We apply a polynomial fitting of order 7 for curve smoothing. The green, blue, and black curves represent the percentage of estimated locations by the NNLEFs in ellipses with CLs 39\%, 66\%, and 95\%, respectively. The dashed colored lines represent the CRBs on location accuracy for the corresponding confidence ellipses (derived in section~\ref{CRB}). From the figure, we observe that the performance for the NNLEF becomes approximately consistent when $P_n\ge$ 8. We do see negligible performance improvement for a few random NNLEFs with $P_n$ in the higher range. These NNELFs are not recommended as they will increase the number of training parameters and computational costs by many folds which do not justify the minimal performance improvement. For example, the number of training parameters for the NNLEF with 8, 30, 40, and 50 neurons in the hidden layer is 66, 242, 322, and 402, respectively.

\begin{figure}[t!]
\centering
\includegraphics[width=0.5\textwidth]{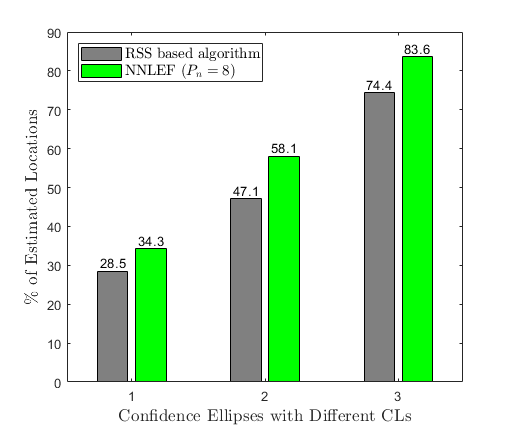}
\caption{Performance comparison for the NNLEF (with $P_n$~=~8) and a state-of-the-art RSS based algorithm. All the simulation parameters are identical to those used in Fig.~\ref{Figure3}. One can see that the NNLEF performs more efficiently when compared to the RSS based algorithm.\label{Figure4}}
\end{figure}

Next, we compare the performance for the NNLEF (with $P_n$ in the recommended range, i.e., 8) with a state-of-the-art RSS based algorithm. The RSS based algorithm minimizes the root mean square error between the measured RSS values in the test and training set to estimate a vehicle's location, i.e.,
\begin{align*}
\hat{\textbf{x}}_e=min\bigg(\frac{1}{N}\sum_{m=1}^{M}(\textbf{r}_{test}-\textbf{r}_{train_m})\bigg),
\end{align*}
where $\textbf{r}_{test}$, and $\textbf{r}_{train}$ are the testing and training set RSS vectors, respectively, and $M$ represents the total number of samples in the training set. In Fig.~\ref{Figure4} we plot the percentage of the estimated test locations in each confidence ellipse.

\begin{figure}[t!]
\centering
\includegraphics[width=0.5\textwidth]{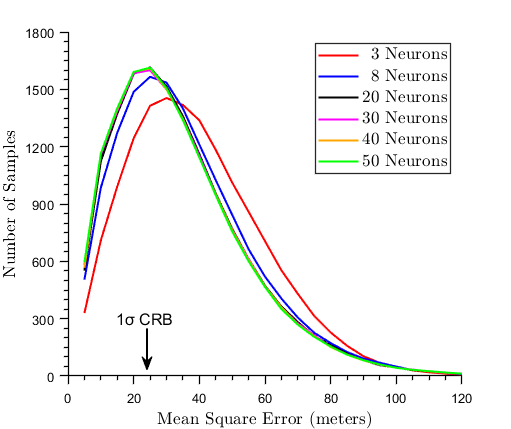}
\caption{Performance evaluation for the NNLEF with $P_n$~=~8 (in the recommended range) and other NNLEFs with random $P_n$. Here, we use similar parameter settings as used in Fig.~\ref{Figure3} but a different performance evaluation metric, i.e., mean square error (as in~\cite{kumar2018feed}). We see that NNLEF with $P_n$~=~3 is under performing. Moreover, we observe an equal performance for the NNLEF with $P_n$~=~8, and the other NNLEFs with higher $P_n$.\label{Figure6}}
\end{figure}

In order to further validate the performance for the NNLEF with the derived architecture, i.e., $P_n=$ 8, in comparison to other NNLEFs with random architectures, i.e., with $P_n=$ equal to 3, 20, 30, 40, and 50, we use a different performance metric as in~\cite{kumar2018feed}, i.e., mean square error (MSE), which is defined as MSE~=~$\sqrt{{(\hat{x}_e-x_0)}^2+{(\hat{y}_e-y_0)}^2}$. Using the same parameter settings as used in Fig.~\ref{Figure3}, we plot the MSE for all the NNLEFs in Fig.~\ref{Figure6}. The x-axis indicates the MSE bin spacing in tens of meters, while the y-axis shows the number of samples in each MSE bin. The solid arrow pointing at the x-axis is the 1$\sigma$ CRB in meters. We see an equivalent performance for all the frameworks. This highlights the fact that a high $P_n$ does not always relate to the NNLEF's performance improvement.

\subsection{Analysis Using Real-world Data}\label{SD}

We now present numerical results by taking into account real-world RSS measurements. These measurements have a multipath factor (from the ground) and noise elements in them. The RSS measurements from random vehicles were collected in a 150 X 150 square meters area by 3 RSUs (installed at (0m, 0m), (-25.5m, 47.6m), and (-8.6m, -46.9m)). 3 devices were used to mimic 3 RSUs. Each device independently measured RSS from the random vehicles at a frequency of 1 RSS measurement per second. Slowly moving Wi-Fi modems (802.11g) with a single antenna (at the same height as RSUs antennas), and a transmission frequency of 2.437 MHz were used to represent slow moving vehicles. These vehicles, equipped with GPS units, reported their GPS locations to the RSUs every second. The RSS measurements at the individual RSUs and the GPS locations of the vehicles were combined utilizing the time stamps (available with both the RSS measurements and the vehicles' GPS locations).

\begin{figure}[t!]
\centering
\includegraphics[width=0.5\textwidth]{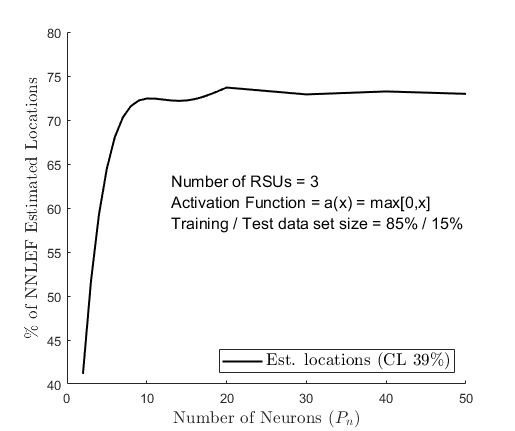}
\caption{The percentage of estimated real-world test data locations by NNLEFs' (with changing $P_n$) in the ellipse with CL~=~95\%.}\label{Figure8}
\end{figure}

At the end of the measurement campaign, the RSS measurement data was thoroughly randomized and divided into a training set (with 85\% of the measurement data) and a test set (of the remaining 15\% measurement data). The training set had the location information of the vehicles while the test set had no such information included. All the NNLEFs (with different $P_n$) were trained using the training set. The trained NNLEFs' were then used to estimate the locations of the vehicles in the test set. In Fig.~\ref{Figure8}, we plot the percentage of the estimated test set locations for NNLEFs with changing $P_n$ in the ellipse with 95\% CL (with polynomial fitting applied). From the figure we observe that performance for the NNLEFs becomes asymptotic once $P_n$ exceeds 8. This validates our earlier claim that a higher $P_n$ may not add much to the performance of the NNLEF, rather it would result in an increase in the computational overheads.
\begin{figure}[t!]
\centering
\includegraphics[width=0.5\textwidth]{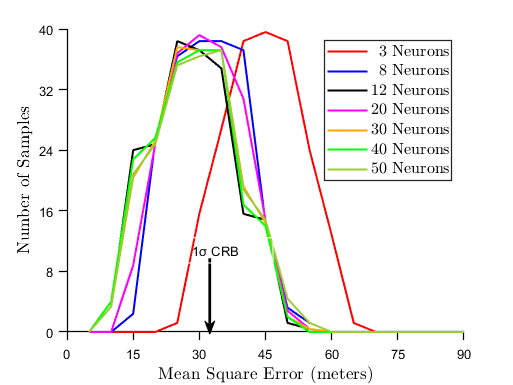}
\caption{MSE performance evaluation for NNLEFs with $P_n$ in the recommended range (i.e., $P_n$~=~7 to 12) and other NNLEFs' with random $P_n$. We observe a poor performance for NNLEF with $P_n$~=~3. On the other hand, we see an equivalent performance for the NNLEFs with the recommended $P_n$ when compared to the other NNLEFs with higher $P_n$.}\label{Figure9}
\end{figure}
We, next compare the performance of an NNLEF with $P_n$~=~8 and 12 (in line with our recommended range for $P_n$) and other NNLEFs with random $P_n$ using MSE metric in Fig.~\ref{Figure9}. We notice that NNLEF with too low a $P_n$, i.e., 3, performs poorly. We also see nearly equivalent performance for the NNLEF (with $P_n$~=~8 and 12 neurons) when compared to the other NNLEFs (with high $P_n$). Here, we observe that NNLEF with $P_n$~=~12 is performing slightly better than NNLEF with $P_n$~=~8.

\section{Conclusion}\label{Conclusion}

In this work we have shown, for the first time, how information-theoretic constructs can be used to decide the number of  hidden-layer  neurons within neural-network architectures for wireless location estimation. Our analysis is confirmed with both simulated data and real-world-data. Our work provides insight into pragmatic architecture design for a wide range of neural-network frameworks beyond location estimation.

\section{Acknowledgment}

The authors acknowledge support by the University of New South Wales, Australia, and Macquarie University, Australia.
Ullah Ihsan acknowledges financial support from the Australian Government through its Research Training Program.

\bibliographystyle{IEEEtran}
\bibliography{ICC}

\end{document}